\documentclass[12pt]{article} 
\usepackage{graphicx,floatflt,amssymb,rotate}
\usepackage{mathrsfs}
\usepackage{amssymb}
\usepackage{amsmath}
\usepackage{color}
\usepackage{multirow}
\usepackage{pstricks}
\usepackage{subfig}
\usepackage{graphicx}
\usepackage{float}
\usepackage[toc,page]{appendix}
\usepackage{amsmath}
\usepackage{hyperref}

\begin{document}

\vskip 30pt  
 
\begin{center}  
{\Large \bf Status of exclusion limits of the  KK-parity non-conserving resonance production  with updated 13 TeV LHC
} \\
\vspace*{1cm}  
\renewcommand{\thefootnote}{\fnsymbol{footnote}}  
{ 
{\sf Avirup Shaw\footnote{email: avirup.cu@gmail.com}} 
} \\  
\vspace{10pt}  
{\small {\em Department of Theoretical Physics, Indian Association for the Cultivation of Science,\\
2A $\&$ 2B Raja S.C. Mullick Road, Jadavpur, Kolkata 700 032, India}}\\ 
   
\normalsize

\end{center}

\begin{abstract}  
\noindent

Unequal strengths of boundary localised terms lead to non-conservation of the Kaluza-Klein (KK) parity in the 4 + 1 Universal Extra Dimensional model. Consequently the first excited KK-partners of Standard Model particles are not stable by any symmetry. In this article using the latest 13 TeV Large Hadron Collider (LHC) results, we revisit the resonant production of first KK-excitations of the neutral gauge bosons ($G^1$, $B^1\ \textrm{and} \ W_3^1$) and their subsequent decay. {\it Specifically $G^1$ (first KK-excitation of gluon) decays to $t\bar{t}$ pair and $B^1/W_3^1$ (first KK-excitation of electroweak gauge bosons) decay to $\ell^{+}\ell^{-} (\ell \equiv e, \mu)$ pair}. We find the exclusion limits of model parameters obtained at 95\% C.L. from the non-observation of these channels have been shifted towards the lower side of the parameter space compared to our previous analysis at 8 TeV.

\vskip 5pt \noindent  
\texttt{PACS Nos:~11.10.Kk, 12.10.Dm, 12.60.−i } \\  
\texttt{Key Words:~Kaluza-Klein, Resonance production, LHC}  
\end{abstract}

\renewcommand{\thesection}{\Roman{section}}  
\setcounter{footnote}{0}  
\renewcommand{\thefootnote}{\arabic{footnote}}  

\section{Introduction}

After a long anticipation LHC at Run-I has been able to discover only the missing piece called Higgs boson of Standard Model (SM) \cite{atlas, cms}. At the same time it is unable to resolve any puzzle among the different long standing issues (Dark Matter, neutrino mass and mixing, gauge hierarchy, CP violation, etc.,) of SM. Even any new physics beyond SM (BSM) has not been detected. In the present days the LHC is now running at $\sqrt{S}=13$ TeV. The prime goal of Run-II is to look for BSM signature. However, both the collaborations ATLAS and CMS have only reported some small local excesses \cite{atlas_excess, cms_excess} over the SM predictions, which need to be verified by thorough analysis at Run-II. At this moment, it would be very relevant job to study or revisit the exclusion limits of the existing BSM physics. {\it In literature we have found several examples where the various collaborating groups have updated their existing results with new data. For example, in the context of minimal supersymmetric standard model (MSSM), there are two references (PRD {\bf 88} (2013) 035011 and PRD {\bf 93} (2016) 075004 ) where the authors (B. Bhattacherjee et al) have studied the status of 98 GeV Higgs boson with different versions of LHC data.} With this spirit in Universal Extra Dimensional (UED) model, one of the popular incarnations of BSM physics, we re-examine the exclusion limits achieved via non-observation of resonance production driven by KK-parity-non-conserving interactions in the light of 13 TeV LHC data \cite{Sirunyan:2017uhk, Aaboud:2017buh}.

The UED model \cite{acd} is characterised by an extra space-like dimension $y$ which is flat and compactified on a circle $S^1$ of radius $R$. All the SM particles can access the dimension $y$. From the four dimensional (4-D) point of view, each of the SM particles has  infinite towers of KK-modes specified by an integer $n$, called the KK-number. $n=0$ modes being labeled as the SM particles. KK-number of a particle is a measure of its momentum along the fifth direction ($y$). One might expect KK-number to be a conserved quantity by the virtue of extra dimensional momentum conservation. However, a $Z_2$ symmetry ($y \rightarrow-y$) has been imposed to incorporate SM chiral fermions. Consequently the translational symmetry along the extra dimension is destroyed and and we encounter KK-number non-conserving interactions. The compactified space is now called $S^1/Z_2$ orbifold which extends only from $y = 0$ to $y = \pi R$. After orbifolding, however, a subgroup of KK-number conservation known as KK-parity\footnote{This KK-parity is equivalent to reflection symmetry of the action with respect to the line $y = \frac{\pi R}{2}$.} can still remain a symmetry of 4-D action.  KK-parity of a state (labeled by KK-number $n$) is defined as $(-1)^n$. This parity ensures the stability of lightest ($n = 1$) KK-particle (LKP) which can be treated as potential Dark Matter candidate in this scenario. Phenomenology of this UED model from the different perspective can be found in the literature \cite{nath}-\cite{flacke2}. 

The effective mass profile of a particle at $nth$ KK-level is $\sqrt{m^2+(nR^{-1})^2}$, $m$ being the corresponding SM mass which is very much lower than $R^{-1}$. Therefore, this model suffers from a deficiency due to the degenerate mass spectrum. Nevertheless, this mass degeneracy could be avoided by radiative corrections. There are two types of radiative corrections, one is called finite bulk correction, and while the other is called boundary correction depending on logarithmic value of cut-off\footnote{Since UED is an extra dimensional theory and hence should be treated as an effective theory valid up to cut-off scale $\Lambda$.} scale $\Lambda$. So it will be relevant to incorporate 4-D kinetic, mass and other necessary interaction terms for the KK-states at the two special points ($y = 0$ and $y = \pi R$) of  $S^1/Z_2$ orbifold. Because these terms can be treated as necessary counterterms for cut-off dependent loop-induced contributions \cite{georgi, cms1, cms2} of the five dimensional (5-D) theory. A very special assumption has been taken in the minimal UED (mUED) models such that the loop-induced contributions exactly vanish at the cut-off scale
$\Lambda$. However, this unique simplification can be discarded. Even without performing the actual radiative corrections one might consider kinetic, mass as well as other interaction terms localised at the fixed boundary points ($y = 0$ and $y = \pi R$) to parametrise these unknown corrections. Hence this scenario is known as non-minimal UED (nmUED) \cite{Dvali}-\cite{ddrs1}. Coefficients of different boundary localised terms (BLTs) as well as the radius of compactification ($R$) can be identified as free parameters of this model. One can constrain these parameters using several experimental inputs. In literature, one finds the bounds on the values of BLT parameters from the consideration of electroweak observables \cite{flacke}, $S$, $T$ and $U$ parameters \cite{delAguila_STU, flacke_STU, bmm}, relic abundance \cite{tommy, ddrs2}, SM Higgs boson production and its decay \cite{tirtha}, study of LHC experiments \cite{asesh, drs, as}, $R_b$ \cite{zbb}, branching fraction of  $B_s \rightarrow \mu^+ \mu^-$ \cite{bmm} and $B \rightarrow X_s \gamma$ \cite{bsg}, flavour violating rare top decay \cite{Dey:2016cve}, $\mathcal{R}(D^{(*)})$ \cite{Biswas:2017vhc} and Unitarity of scattering amplitudes involving KK-excitations \cite{Jha:2016sre}.

In this article we create non-conservation of KK-parity\footnote{This can be viewed as R-parity non-conservation in supersymmetry.} by adding boundary localised terms of unequal strengths \cite{ddrs1, drs, as}. Consequently $n = 1$ KK-states are no longer stable and decay to pair of SM particles. Capitalising on this KK-parity-non-conserving coupling, we revisit the resonant production of KK-excitations of the neutral gauge bosons at the LHC and their
subsequent decay into SM fermion pair. We use the latest LHC data \cite{Sirunyan:2017uhk} and \cite{Aaboud:2017buh} of CMS and ATLAS Collaborations for search of resonant high-mass new phenomena in top-antitop quark pairs ($t\bar{t}$) and dilepton ($\ell^{+}\ell^{-}$) final states respectively. From this analysis we will put constraints on the BLT parameters as well as on the size of the radius of compactification.

The plan of this article is as follows. First, we discuss the relevant couplings and masses in the framework of UED with unequal strength of BLT parameters. We then revisit the exclusion limits obtained via $t\bar{t}$ signal from $G^1$ production and $\ell^{+}\ell^{-}$ signal from the combined production of the $B^1\ \textrm{and} \ W_3^1$ at the 13 TeV LHC. Furthermore, we compare the limits obtained from the current analysis with that our previous 8 TeV analyses \cite{drs, as}. Subsequently we will explore the reasons for which we will obtain the deviation of the limits (in each case for both signal) at 13 TeV LHC. Finally we will summarise the results in section V.

\section{A glimpse of KK-parity-non-conserving nmUED}

In this section we discuss the primitive terminologies of KK-parity non-conserving nmUED \cite{ddrs1, drs, as}. For the purpose of detailed analysis of the model we refer \cite{Dvali}-\cite{ddrs1}. In the following action we choose the  unequal strengths of the boundary terms at the two boundary points ($y = 0$ and $y = \pi R$). Consequently the KK-parity will not be conserved anymore. However, if the strengths of the boundary terms be equal then the KK-parity will be restored and we can have the potential Dark Matter candidate \cite{ddrs2}.

Let us begin with the action of 5-D fermionic fields $\Psi_{L,R}$ including boundary localised kinetic terms (BLKTs) \cite{schwinn, ddrs1, drs, as}: 
\begin{eqnarray} 
S_{fermion} = \int d^5x \left[ \bar{\Psi}_L i \Gamma^M D_M \Psi_L 
+ \{r^a_f\delta(y)+r^b_f\delta(y - \pi R)\} \bar{\Psi}_L i \gamma^\mu D_\mu P_L\Psi_L  
\right. \nonumber \\
\left. + \bar{\Psi}_R i \Gamma^M D_M \Psi_R
+ \{r^a_f\delta(y)+r^b_f\delta(y - \pi R)\}\bar{\Psi}_R i \gamma^\mu D_\mu P_R\Psi_R
\right].
\label{faction}
\end{eqnarray} 
$M=(0, 1\ldots 4)$. $r^a_f, r^b_f$ are the coefficients\footnote{To respect the chiral symmetry we have chosen equal strengths of BLKTs for both fermion fields $\Psi_{L,R}$.} of the BLKTs localised at the two fixed points ($y = 0$ and $y = \pi R$). We can decompose the 5-D four component fermion fields $\Psi_{L,R}$ into two component chiral spinors using the following relations \cite{schwinn, ddrs1, drs, as}:

\begin{equation} 
\Psi_L(x,y) = \begin{pmatrix}\phi_L(x,y) \\ \chi_L(x,y)\end{pmatrix}
=   \sum_n \begin{pmatrix}\phi^{(n)}_L(x) f_L^n(y) \\ \chi^{(n)}_L(x) g_L^n(y)\end{pmatrix}, 
\label{fiveDL}
\end{equation}
\begin{equation} 
\Psi_R(x,y) = \begin{pmatrix}\phi_R(x,y) \\ \chi_R(x,y) \end{pmatrix} 
=   \sum_n \begin{pmatrix}\phi^{(n)}_R(x) f_R^n(y) \\ \chi^{(n)}_R(x) g_R^n(y) \end{pmatrix}. 
\label{fiveDR} 
\end{equation}
Applying suitable boundary conditions \cite{carena, ddrs1}, we can have the following KK-wave-functions which are simply denoted by $f$ for illustrative purposes \cite{carena, ddrs1, drs, as}:
\begin{eqnarray}
f^n(y) &=& N_n \left[ \cos (m_n y) - \frac{r_f^a m_n}{2} \sin (m_n
y) \right] \;,\;\;  0 \leq y < \pi R,   \nonumber \\ 
f^n(y) &=& N_n \left[ \cos (m_n y) + \frac{r_f^a m_n}{2} \sin (m_n
y) \right] \;,\;\; -\pi R \leq y < 0.
\label{sol1}
\end{eqnarray}

KK-masses $m_n$ for  $n = 0,1, \ldots$ satisfy the following transcendental equation \cite{carena, ddrs1, drs, as}:
\begin{equation} 
(r^a_f r^b_f ~m_n^2 - 4) \tan(m_n \pi R)= 2(r^a_f + r^b_f) m_n \;.
\label{trans1}
\end{equation}

For the purpose of illustration, we have chosen two different strategies to study the KK-parity-non-conservation. In the first case, we demand that  the strength of BLKTs (at two fixed point $y = 0$ and $y = \pi R$) are equal for fermions, i.e., $r^a_f = r^b_f \equiv r_f$, while for the second case BLKTs vanish at one of the fixed boundary points. For the later case we choose, $r^a_f \neq 0, ~r^b_f = 0$ and in this situation Eq.\;(\ref{trans1}) reduces to \cite{carena, ddrs1, drs, as}:
\begin{equation} 
\tan(m_n \pi R)=-\frac{r^a_f m_n}{2} .
\label{trans3f}
\end{equation}

$N_n$ is the normalisation constant for $nth$ KK-mode and obtained from
orthonormality condition \cite{carena, ddrs1, drs, as}:
\begin{equation}
\int_0 ^{\pi R} dy \left[1 + r^a_f \delta(y) + r^b_f \delta(y - \pi R)
\right] ~f^n(y) ~f^m(y) = \delta^{n m}.
\label{norm}
\end{equation} 

For the first case
\begin{equation}
 N_n = \sqrt{\frac{2}{\pi R}}\left[ \frac{1}{\sqrt{1 + \frac{r_f^2 m_n^2}{4} 
+ \frac{r_f}{\pi R}}}\right],
\label{norm1}
\end{equation}
when strength of boundary terms are equal i.e., $r^b_f = r^a_f\equiv r_f$. And for the second situation when $r^b_f = 0$ and we set $r^a_f \equiv r_f$, one finds:
\begin{equation}
 N_n = \sqrt{\frac{2}{\pi R}}\left[ \frac{1}{\sqrt{1 + \frac{r_f^2 m_n^2}{4} 
+ \frac{r_f}{2 \pi R}}}\right].
\label{norm2}
\end{equation}

To this end let us discuss the values of BLT parameters which we would like to use in our analysis. If $\frac{r^a_f}{R}$ $(<<1)$ the KK-mass formula approximately reduces to (using Eq. \ref{trans3f}) \cite{as}:
\begin{equation}
{ 
m_n\approx\frac{n}{R}\left(\frac{1}{1+\frac{r^a_f}{2\pi R}}\right)\approx\frac{n}{R}\left(1-\frac{r^a_f}{2\pi R}\right).}
\label{apprx}
\end{equation} 
It is evident from the above expression that the KK-mass reduces with increasing positive values of $r^a_f$. This feature also holds good when the BLKTs are non-vanishing at both the boundary points.

Now it is clear from the Eqs.\;\ref{norm1} and \ref{norm2}  that, for  $\frac{r_f}{R}<-\pi$~(when BLKTs are present at the two boundary points)~and $\frac{r_f}{R}<-2\pi$ (when BLKTs are present only at one of the two boundary points) the squared norm of zero mode solutions become negative. Furthermore, for $\frac{r_f}{R}=-\pi$~(when BLKTs are present at the two boundary points)~and $\frac{r_f}{R}=-2\pi$ (when BLKTs are present only at one of the two boundary points) the solutions turn out to be divergent. Beyond these limits the fields behave like ghost fields and consequently the values of $\frac{r_f}{R}$ beyond these region should be discarded. Also analysis of electroweak precision data reveals that the negative values of BLT parameters are not so competitive \cite{bmm}. Furthermore, negative values of BLT parameters are less attractive due to phase space consideration as negative values of BLT parameters enhance the KK-mass which in turn suppresses the production cross section. Hence we will use the positive values of BLKTs in the rest of our analysis.

Let us turn to the action for the 5-D gauge fields. In presence of BLKTs at the boundary points ($y=0$ and $y=\pi R $) this can be written down as \cite{ddrs1, drs, as}:
\begin{eqnarray}
S_V &=& -\frac{1}{4}\int d^5x \bigg[ G^\lambda_{MN} G^{\lambda MN}+   \left\{r^a_G \delta(y) + r^b_G \delta(y - \pi R)\right\} G^\lambda_{\mu\nu} G^{\lambda\mu \nu}\nonumber \\
&&+W^i_{MN} W^{iMN}+   \left\{r^a_W \delta(y) + r^b_W \delta(y - \pi R)\right\} W^i_{\mu\nu} W^{i\mu \nu}\nonumber \\
&&+B_{MN} B^{MN}+ \left\{r^a_B \delta(y) +  r^b_B \delta(y - \pi R)\right\} B_{\mu\nu} B^{\mu \nu}\bigg].
\label{pure-gauge}
\end{eqnarray}

Here, $r^a_V$ and $r^b_V$ ($V \equiv G, W, B$) parametrise the strength of the BLKTs for the gauge fields. 5-D field strength tensors are given below:
\begin{eqnarray}\label{ugfs}
G_{MN}^\lambda &\equiv& (\partial_M G_N^\lambda - \partial_N G_M^\lambda-{\tilde{g}_3}f^{\lambda\rho\sigma}G_M^\rho G_N^\sigma),\\ \nonumber
W_{MN}^i &\equiv& (\partial_M W_N^i - \partial_N W_M^i-{\tilde{g}_2}\epsilon^{ijk}W_M^jW_N^k),\\ \nonumber
B_{MN}&\equiv& (\partial_M B_N - \partial_N B_M).
\end{eqnarray}
$G^\lambda_M$ ($\lambda=1,\ldots 8$), $W^i_M$ ($i=1,2,3$) and $B_M$  are the 5-D gauge fields corresponding to $SU(3)_C$ $SU(2)_L$ and $U(1)_Y$ gauge group respectively. Generically $y$-dependent KK-wave-functions for the gauge fields ($V_M\equiv(V_\mu, V_4)$) can be written in the following way \cite{ddrs1, drs, as}:

\begin{equation}\label{Amu}
V_{\mu}(x,y)=\sum_n V_{\mu}^{(n)}(x) a^n(y)~;~V_{4}(x,y)=\sum_n V_{4}^{(n)}(x) b^n(y).
\end{equation}

A convenient gauge choice\footnote{A general analysis on gauge-fixing action and gauge-fixing mechanism in the nmUED model can be found in \cite{gf}.} for this model would be putting $V_4\rightarrow0$. This gauge choice would easily eliminate the undesirable
terms in which $V_\mu$ couples to $V_4$ via derivative \cite{ddrs1, drs, as}. This is the main purpose of gauge-fixing mechanism. As we are interested in $V_\mu$ and its interactions with other physical particles, setting $V_4\rightarrow0$ is as good as Unitary gauge \cite{ddrs1, drs, as}. 

KK-masses for gauge fields are very similar to the fermions and can be obtained from Eq.\;\ref{trans1} and Eq.\;\ref{trans3f} in a similar manner. The detail discussions on the gauge fields in nmUED model is readily available in \cite{ddrs1}. 

Furthermore, in Ref.\;\cite{gf} we have shown that when the gauge symmetry is spontaneously broken, the BLKT parameters of gauge bosons and Higgs should be equal for the purpose of proper gauge-fixing. This condition leads to the equality of BLKT parameters
of $SU(2)_L$ and $U(1)_Y$ gauge bosons i.e., $(r^a_B=r^a_W\equiv r^a_V)~{\rm and}~(r^b_B=r^b_W\equiv r^b_V)$. Hence KK-masses of $B$ and $W_3$ are equal. However, for the $SU(3)_C$ gauge boson we can independently choose the BLKT parameters ($r^a_G, r^b_G$), which are different from the electroweak sector.

\begin{figure}[H] 
\centering
\subfloat[]{\includegraphics[width=7cm,height=4.2cm]{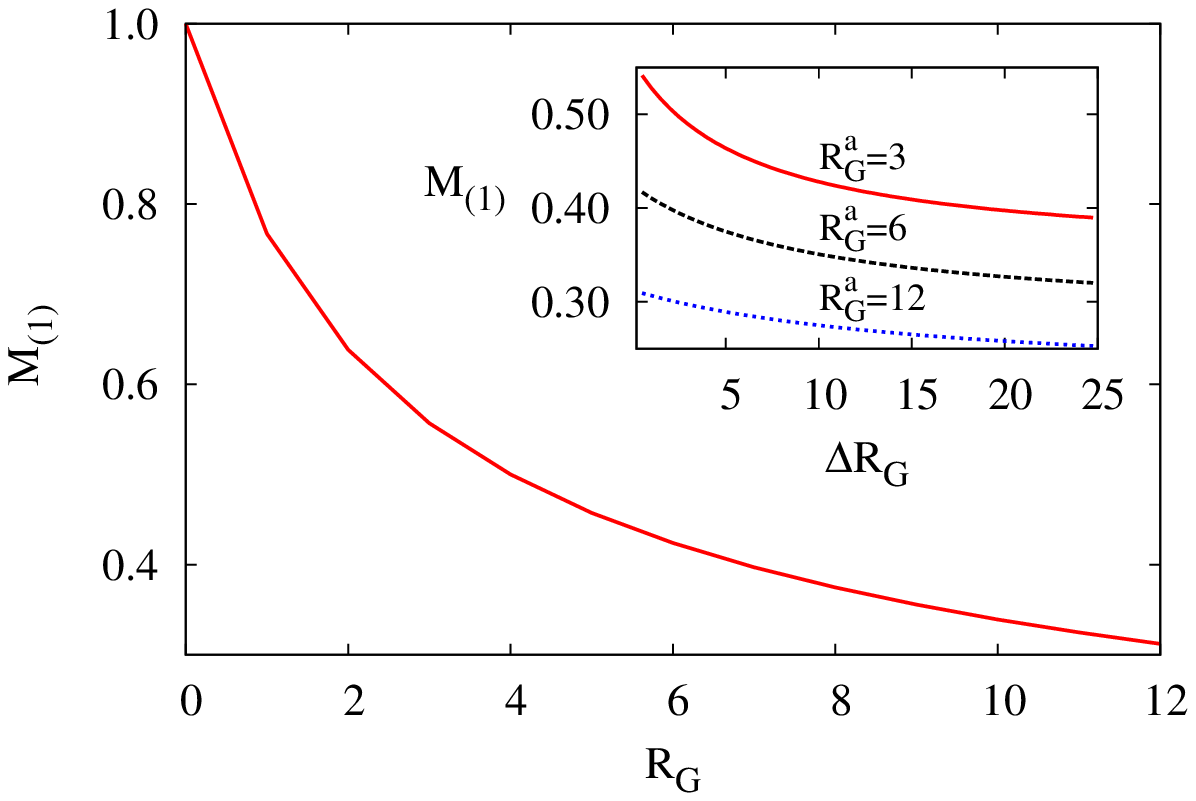}}
\subfloat[]{\includegraphics[width=7cm,height=4.2cm]{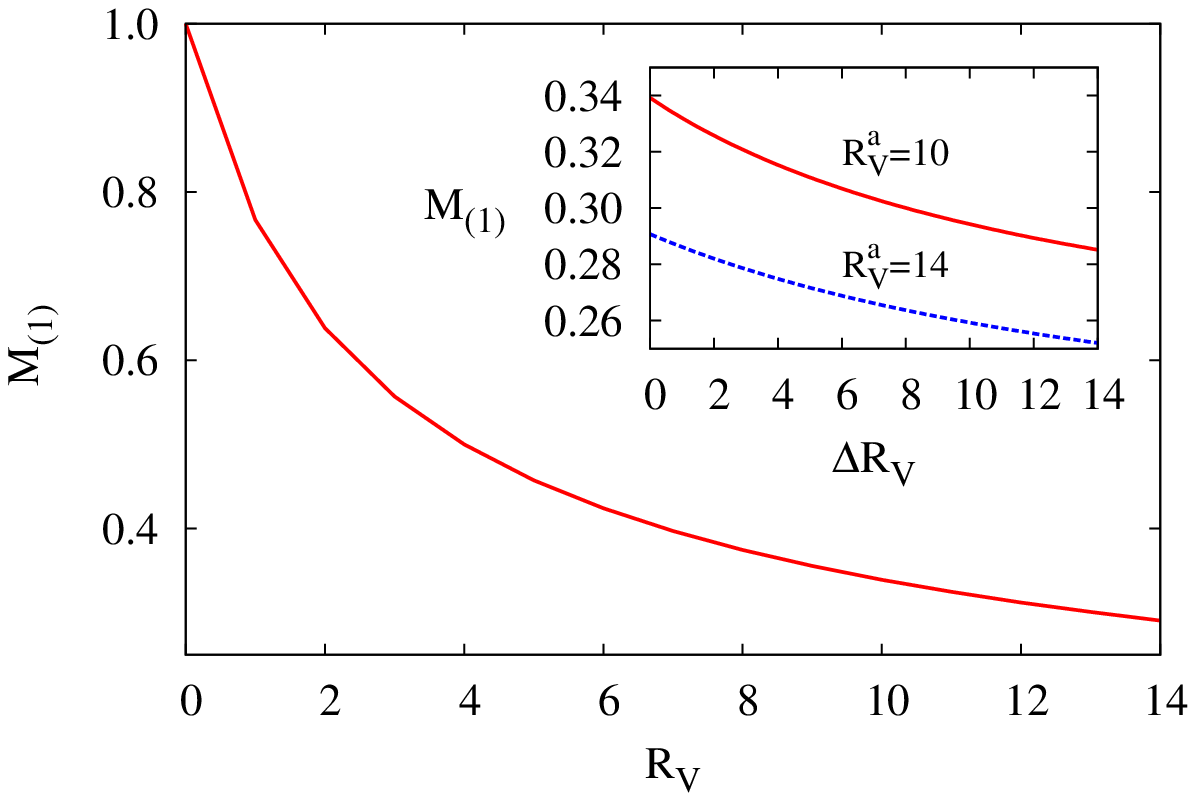}}\\
\subfloat[]{\includegraphics[width=7cm,height=4.2cm]{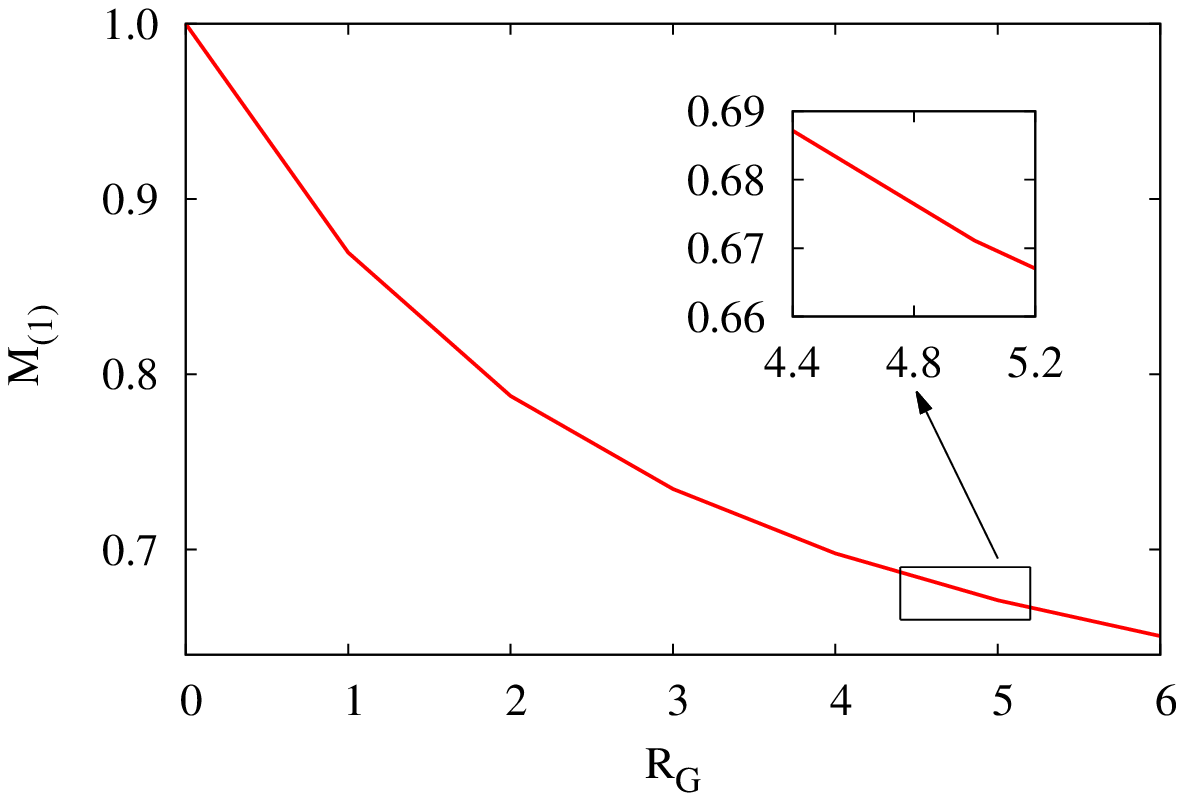}}
\subfloat[]{\includegraphics[width=7cm,height=4.2cm]{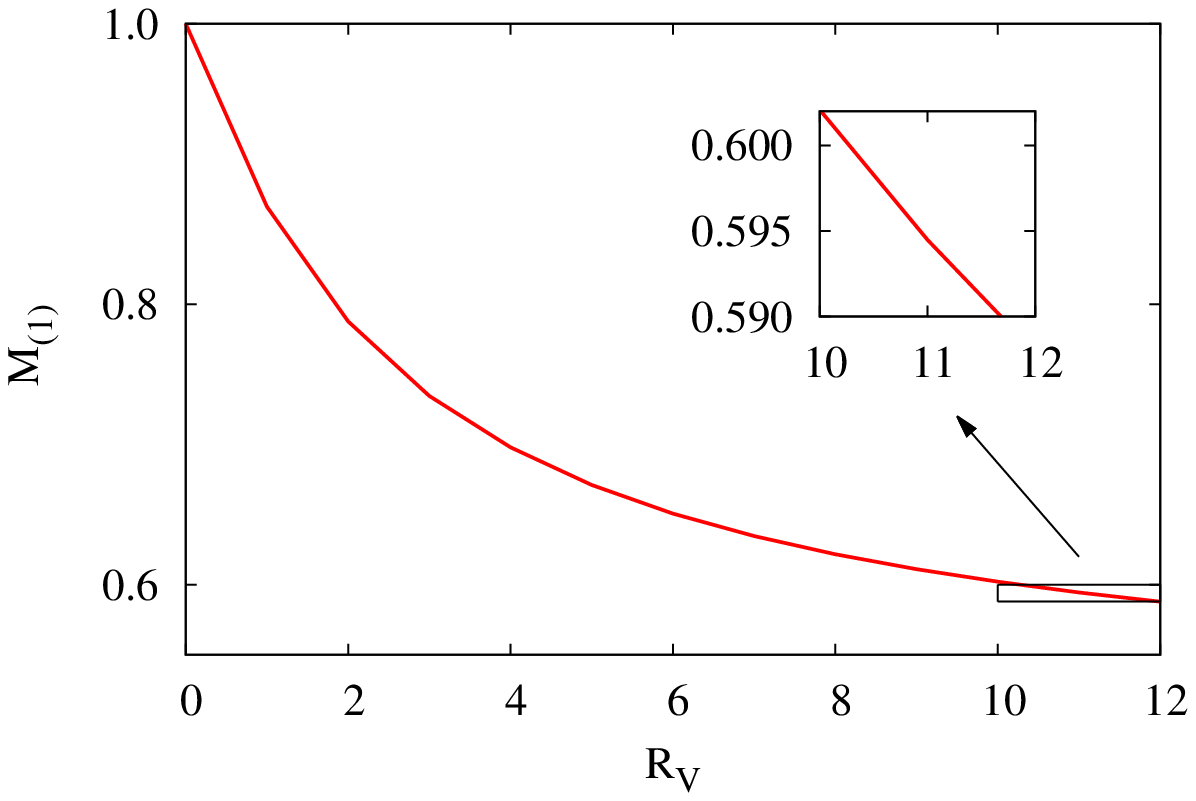}}
\caption{Left side: The upper (a) and lower (c) panels  show the variation of $M_{(1)}(=m_{G^{(1)}} R)$ as a function of BLKT parameters which we will use in $t\bar{t}$ production. Similarly the upper (b) and lower (d) panels of right side show the variation of $M_{(1)}(=m_{V^{(1)}} R)$ corresponding to $\ell^{+} \ell^{-}$ production. For the purpose of detail illustrations one can see the text and also the Refs.\;\cite{ddrs1, drs, as}.}
\label{mass}
\end{figure}

In the different panels ((a), (b), (c) and (d)) of Fig.\;\ref{mass} we have plotted dependence of scaled KK-mass for the first KK-excited gauge fields $G(\equiv {\rm gluon})$ and $V(\equiv W_3, B$) with respect to appropriate ranges of BLKT parameters what we use in our main analysis. However,  the characteristic are similar for all kind of fields. In the {\it upper} panels ((a) and (b)) we have shown the mass profile with respect to $R_\alpha(={r_\alpha} {R^{-1}} $) in the symmetric limit ($R^a_\alpha=R^b_\alpha \equiv R_\alpha$) ($\alpha \equiv G, B, W_3$). In the inset we have  presented the variation of $M_{(1)}(=m_{\alpha^{(1)}} R)$ with asymmetric parameter $\Delta R_\alpha(={(r^b_\alpha-r^a_\alpha)}{R^{-1}})$, for different choices of $R^a_\alpha$. Similarly in the {\it lower} panels ((c) and (d)) we have presented the variation of $M_{(1)}$ with respect to $R_\alpha(={r_\alpha} {R^{-1}}$) when boundary terms are present only at the boundary point $y=0$. Here, also the values of BLKT parameters shown in the inset plots are used in our analysis. The two left panels ((a) and (c)) show the range of BLKT parameter space which we will use in $t\bar{t}$ resonance production where as the two right panels ((b) and (d)) show the range of BLKT parameter space for $\ell^{+} \ell^{-}$ resonance production. In all the cases the KK-masses are decreased with the increasing positive values of BLKTs. The detail characteristic features of these plots can be found in \cite{ddrs1, drs, as}.
 
\section{KK-parity-non-conserving coupling of $V^1(\equiv G^1, B^1$ and $W^1_3$) with zero mode fermions}
Conservation of KK-parity is an inherent property of UED model, even it is still conserved in presence of BLTs of equal strength. However, non-conservation of KK-parity can be generated with unequal strength of BLTs. In our work we originate this non-conservation in two different ways. In one set up we consider the strengths of BLKTs for fermions equal at the boundary points i.e., $r^a_f = r^b_f\equiv r_f$ while for the gauge bosons $r^a_V \neq r^b_V$. In the other option we assume that the BLKTs are present only at the $y = 0$ fixed point for both the fermions and the gauge bosons. Utilising the above alternatives we give rise the interacting coupling between gauge boson at $n=1$ KK-level with pair of SM (zero mode) fermions and is given by \cite{ddrs1, drs, as}:

\begin{eqnarray}
\label{kkvcoplng}
g_{V^{1}f^{0}f^{0}}=\left\{\begin{array}{rl}
                \displaystyle \tilde{g}\int_0 ^{\pi R}
dy \; \left[1+r^a_f\delta(y)+r^b_f\delta(y-\pi R)\right]f_{L}^{0}f_{L}^{0}a^{1},\\
                \displaystyle \tilde{g}\int_0 ^{\pi R}
dy \; \left[1+r^a_f\delta(y)+r^b_f\delta(y-\pi R)\right]g_{R}^{0}g_{R}^{0}a^{1}.
\end{array}\right.
\end{eqnarray}

$\tilde{g}$  represents 5-D gauge coupling which is connected to the conventional 4-D gauge coupling $g$ through the following relation \cite{ddrs1, drs, as}: 
\begin{equation}
\tilde{g} = g ~\sqrt{\pi R \left(1+\frac{r^a_V+ r^b_V}{2\pi R}\right)}~.
\end{equation}
Here  $f_L^0, g_R^0$ are denoted as wave-functions for zero mode fermion and $a^1$ is identified as wave-functions for the first excited state ($n = 1$) of KK-gauge bosons.

In our first choice $y$-dependent wave-functions are given as follows:
\begin{equation}
f_{L}^{0} = g_{R}^{0} = \frac{1}{\sqrt{\pi R(1 + \frac{R_f}{\pi})}}, 
\end{equation}
and
\begin{equation}
a^{1} = N^V_1  
\left[\cos \left( \frac{M_{(1)}y}{R} \right)-\frac{R^a_V M_{(1)}}{2}
\sin \left(\frac{M_{(1)}y}{R}\right)\right],
\end{equation}

with normalisation constant $$
N^V_1 = \sqrt{\frac{1}{\pi R}}~\sqrt{\frac{8(4+M_{(1)}^{2}{R^b}^2_V)}
{2\left(\frac{R^{a}_V+R^{b}_V}{\pi}\right)(4+M_{(1)}^{2}R^{a}_VR^{b}_V)
+(4+M_{(1)}^{2}{R^{a}}^2_V)(4+M_{(1)}^{2} {R^{b}}^2_V)}}~~,$$ 

where, $M_{(1)}=m_{V^{(1)}}R$, $R_f={r_f}{R^{-1}}$, $R^a_V={r^a_V}{R^{-1}}$, and $R^b_V={r^b_V}{R^{-1}}$. Utilising the above finally we acquire the effective 4-D coupling \cite{ddrs1, drs, as}:
\begin{eqnarray}\label{coup1}
g_{V^{1}f^{0}f^{0}}&=&\frac{g\sqrt{\pi R \left(1+\frac{R^a_V+ R^b_V}{2 \pi}\right)}N^V_1}{\left(1+\frac{R_{f}}{\pi}\right)}\Bigg[\frac{\sin(\pi M_{(1)})}
{\pi M_{(1)}}\left\{1-\frac{M_{(1)}^{2}R^a_VR_{f}}{4}\right\} \nonumber \\
&&+\frac{R^a_V}{2\pi}\left\{\cos(\pi M_{(1)})-1\right\} 
{+\frac{R_{f}}{2\pi}\left\{\cos( \pi M_{(1)})+1\right\}\Bigg] }.
\end{eqnarray}
This coupling vanishes for the equality condition $R^a_V=R^b_V$ \cite{ddrs1, drs, as}.

In the second case, the $y$-dependent wave-functions are given by:
\begin{equation} 
f_L^{0} = g_R^{0} = \frac{1}{\sqrt{\pi R(1 + \frac{R_f}{2 \pi})}}, 
\end{equation}
and
\begin{equation}
a^{1} = \sqrt{\frac{2}{\pi R}}~\sqrt\frac{1}{1+\left(\frac{R_V
M_{(1)}}{2}\right)^2+\frac{R_V}{2\pi}}
\left[\cos\left(\frac{M_{(1)}y}{R}\right)-
\frac{R_V M_{(1)}}{2}\sin\left(\frac{M_{(1)} y}{R}\right) 
\right] \;.
\end{equation} 

We obtain for this choice \cite{ddrs1, drs, as}:

\begin{eqnarray}\label{coup2}
g_{V^{1}f^{0}f^{0}} &=&  
\frac{\sqrt{2} ~g\sqrt{\left(1+\frac{R_{V}}{2\pi}\right)} }
{\left(1+\frac{R_{f}}{2\pi}\right) 
\sqrt{1+\left(\frac{R_V M_{(1)}}{2}\right)^2+\frac{R_V}{2\pi}}}  
\left(\frac{R_{f}-R_V}{2\pi}\right).  
\end{eqnarray} 

\begin{figure}[t] 
\centering
\subfloat[]{\includegraphics[scale=0.95,angle=0]{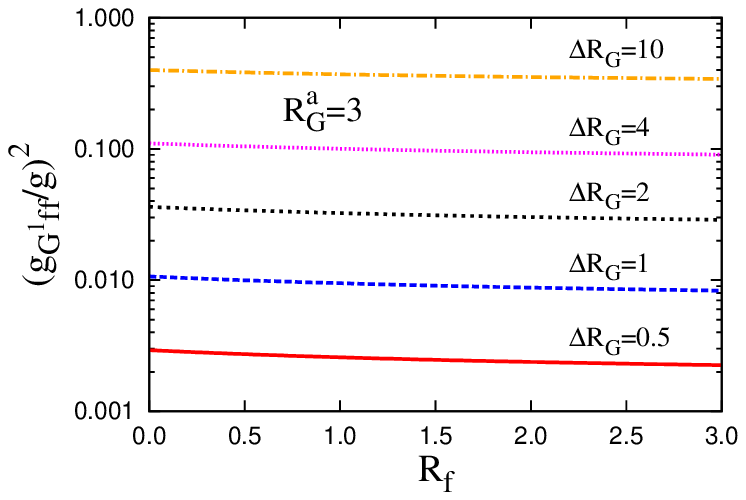}}
\subfloat[]{\includegraphics[scale=0.95,angle=0]{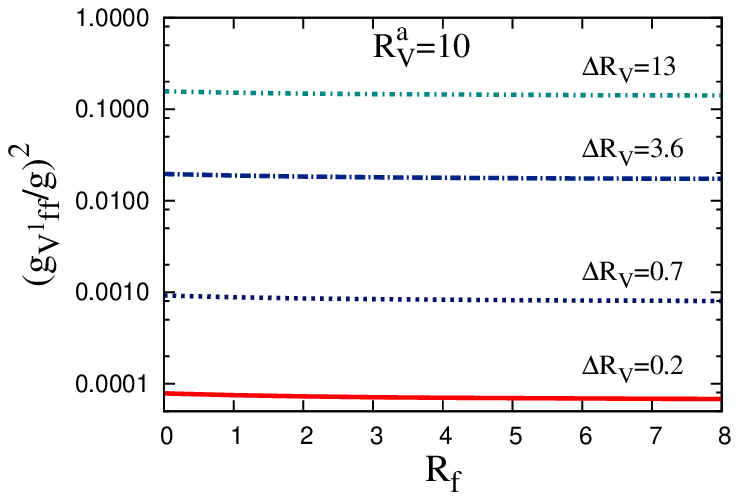}}\\
\subfloat[]{\includegraphics[scale=0.95,angle=0]{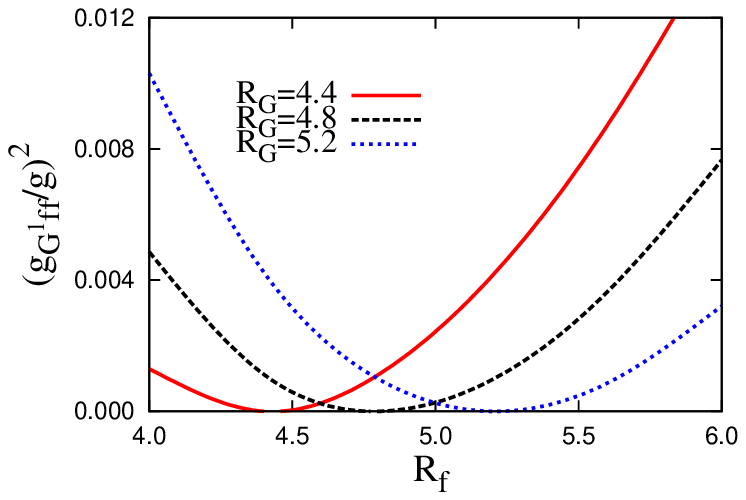}}
\subfloat[]{\includegraphics[scale=0.95,angle=0]{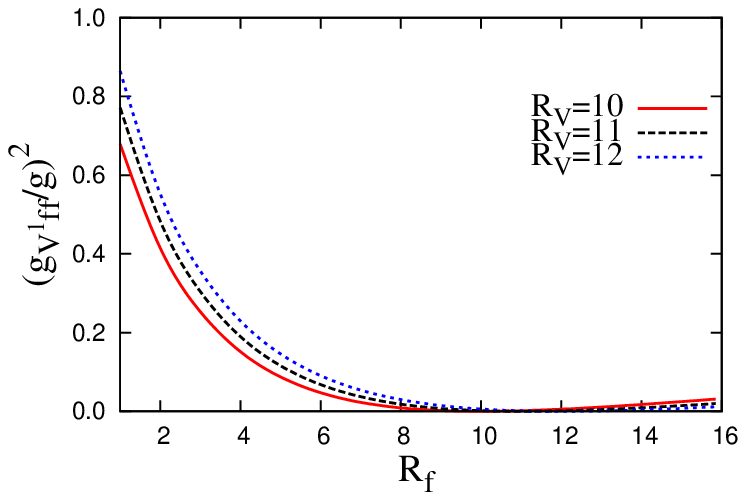}}
\caption{ In the upper (a) and lower (c) panels of left side show the variation of $(g_{G^{1}f^{0}f^{0}}/g)^2$ as a function of BLKT parameters what we use in $t\bar{t}$ production. Similarly the upper (b) and lower (d) panels of right side show the variation of $(g_{V^{1}f^{0}f^{0}}/g)^2$ corresponding to $\ell^{+} \ell^{-}$ production. For the purpose of detail illustrations one can see the text and also the Refs.\;\cite{ddrs1, drs, as}.}
\label{coup_plots} 
\end{figure}

This coupling disappears when $R_f=R_V$ \cite{ddrs1, drs, as}. It can be easily checked from the lower panels of Fig.\;\ref{coup_plots}. 

In Fig.\;\ref{coup_plots} we have shown the variation of scaled KK-parity-non-conserving coupling which can be termed as overlap integrals\footnote{Effective interactions in this model can be achieved by integrating out the 5-D action over the extra space-like dimension after replacing the appropriate $y$-dependent KK-wave-function for the respective fields in 5-D action, see Eq.\;\ref{kkvcoplng}. Consequently some of the interactions are modified by some multiplicative factors which are called
overlap integrals.} for different cases. A careful look at the upper panels ((a) and (b)) reveal that the KK-parity-non-conserving coupling rises with increasing values of asymmetric parameter $\Delta R_\alpha=(R^b_\alpha - R^a_\alpha)$ ($\alpha \equiv G, B, W_3$) while decreases with the increasing values of $R^a_\alpha$. The magnitude of KK-parity-non-conserving coupling decreases mildly with the greater values of $R_f$. In the lower panels ((c) and (d)) the coupling shows oscillatory nature with $R_f$. However, in the region of our interest ($R_V(R_G) > R_f$), the KK-parity-non-conserving coupling rises with higher values of $R_G$ ($R_V$). We have shown the values of the coupling strengths with BLKT parameters which we use in our analysis. The detailed dependence of this coupling strength with respect to the BLKT parameters can be found in Refs.\;\cite{ddrs1, drs, as}.

\section{Gauge boson $V^1(\equiv G^1, B^1$ and $W^1_3$) production and decay}
Up to this point we have all the required ingredients to discuss some phenomenological signals of nmUED.  Specifically at the LHC we are interested to investigate the resonant production of $pp\rightarrow V^{1}$ followed by the decay $V^{1} \rightarrow f^{0}\bar{f}^{0}$, where $f^{0}$ being the SM quarks or leptons\footnote{From now and onwards we will not use superscript \textquotedblleft 0\textquotedblright~ for SM particles.}. Both the production and the decay of $V^{1}$ are governed by KK-parity non-conserving 
couplings which vanish if the strengths of BLT parameters at two boundary points are the same \cite{ddrs1, drs, as}. A compact expression for the production cross section in $pp$ collisions can be written as: 
\begin{equation}\label{int_sig}
\sigma (p p \rightarrow V^{1} + X) = \frac{4 \pi^2 N_c}{3 m_{V^{(1)}}^3}\;\sum_i 
\Gamma(V^{1} \rightarrow q_i \bar q_i)\tau
\int_\tau ^1 \frac{dx}{x}\
\left[f_{\frac{q_{i}}{p}}(x,m_{V^{(1)}}^2) 
f_{\frac{{\bar q_{i}}}{p}}(\tau/x,m_{V^{(1)}}^2) + 
q_i \leftrightarrow \bar q_i \right].
\end{equation} 
Here, $q_i$ and $\bar{q_i}$ represent a generic quark and its antiquark of the $ith$ flavour respectively. Quark (antiquark) distribution function within a proton is represented by $f_{\frac{q_{i}}{p}}$ ($f_{\frac{{\bar q_{i}}}{p}}$). We denote $\tau={m_{V^{(1)}}^2 / S}$, where $\sqrt{S}$
is the $pp$ centre of momentum energy and $N_c$ is the colour factor.
$\Gamma(V^{1} \rightarrow q_i \bar q_i)$ is the decay width of $V^1$ into the SM quark-antiquark pair ($q_i \bar{q_i}$). In the case of  $pp\rightarrow G^{1}\rightarrow t\bar{t}$ channel we have not considered higher order perturbative QCD corrections in our analysis as QCD corrections usually increase cross sections. In this regard our results are probably conservative. 

Decay width of $G^1$ ($n=1$ KK-excitation of {\it gluon}) into $q_i \bar{q_i}$ pair is given by $\Gamma = \left[\frac{{g^{2}_{V^{1}q q}}}{\pi}\right]m_{G^{(1)}}$. In the case of $B^1$ ($n=1$ KK-excitation of $U(1)_Y$ gauge boson) one has $\Gamma = \left[\frac{{g^{2}_{V^{1}q q}}}{32\pi}\right]\left[(Y^{q}_L)^2 + (Y^{q}_R)^2 \right]m_{B^{(1)}}$ (with $Y^q _L$ and $Y^q _R$ being the weak-hypercharges
for the left- and right-chiral quarks), while the decay width of $W^1 _3$ ($n=1$ KK-excitation of neutral $SU(2)_L$ gauge boson) is given by $\Gamma = \left[\frac{{g^{2}_{V^{1}q q}}}{32\pi}\right]m_{W^{(1)}_3}$.
$g^{}_{V^1 q q}$ represents the KK-parity-non-conserving coupling between $q_i \bar{q_i}$ pair and $V^1$ as given in Eqs.\;\ref{coup1} and \ref{coup2}. In the expression of cross section (see Eq.\;\ref{int_sig}), $m_{V^{(1)}}$ denotes the mass eigenvalue of the gauge
boson of $n=1$ KK-excitation. The KK-modes of electroweak gauge bosons ($B$ and $W_3$) also acquire a contribution
to their masses from spontaneous breaking of the electroweak symmetry, however that contribution is not taken into account in our analysis as they are negligible in comparison to extra dimensional contribution. Hence $B^1$ and $W^1_3$ share the same mass eigenvalue.

To determine the numerical values of the cross sections, we use a parton level Monte Carlo code  with parton distribution functions as parametrized in CTEQ6L \cite{CTEQ}.  In this analysis the $pp$ centre of momentum energy is 13 TeV.  Further, we set factorisation scales (in the parton distributions) and renormalisation scale for $\alpha_s$ at $m_{V^{(1)}}$.

We have contrasted the above outcome with two different results from the LHC. In the
following sections we will be going to present the search of $t\bar{t}$ and $\ell^{+} \ell^{-}$ resonances at the LHC running at 13 TeV $pp$ centre of momentum energy. We assume that $G^1~{\rm or}~B^1 (W^1_3)$ to be the lightest KK-particles (LKP) in the respective cases. In this situation after the production of $G^1$ or $B^1 (W^1_3$) (first excited KK gauge boson), the KK-parity conserving decays being kinematically disallowed, the $G^1$ or $B^1 (W^1_3$) decays to a pair of zero-mode fermions (quarks or charged leptons) via the same KK-parity non-conserving coupling. From the lack of observation of such  signals at 95\% C.L., upper bounds have been established on the cross section times  branching fraction\footnote{The branching fraction of $G^1$ to $t\bar{t}$ is nearly $\frac16$ and for $B^1\ (W_3^1)$ to $e^{+}e^{-}$ and $\mu^{+}\mu^{-}$ is nearly $\frac{30}{103}$ ($\frac{2}{21}$).} as a function of the mass of a $t
\bar{t}$ and/or $\ell^{+}\ell^{-}$ resonance. Comparing these bounds with the theoretical predictions in the KK-parity-non-conserving framework one can constrain the parameter space of this nmUED model. To acquire the most up-to-date
bounds we use the latest 13 TeV results from CMS \cite{Sirunyan:2017uhk} and ATLAS \cite{Aaboud:2017buh} data for $t\bar{t}$ and $\ell^{+}\ell^{-}$ resonance production respectively. Results for two different signals for two distinct cases, either BLKTs are non-vanishing at both boundary points or only at one of the two, will be presented in following two sections.

\section{\ {\boldmath$t\bar{t}$} resonance search at 13 TeV and comparative study with 8 TeV analysis} 
\begin{figure}[t]
\begin{center}
\includegraphics[scale=1,angle=0]{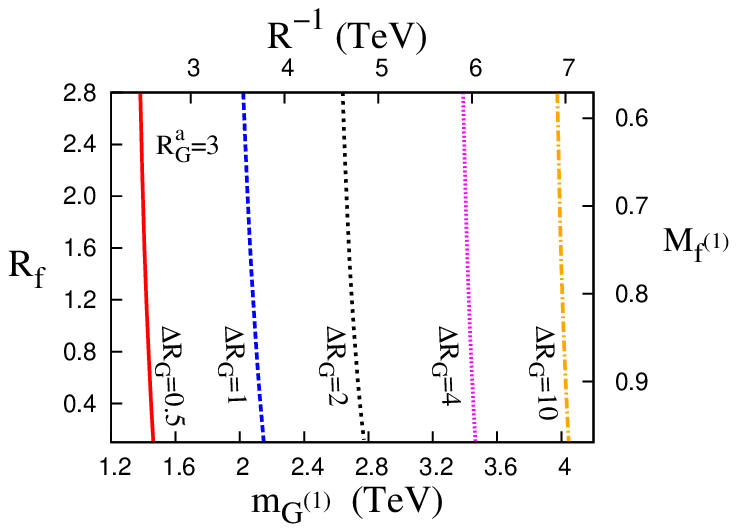}
\includegraphics[scale=1,angle=0]{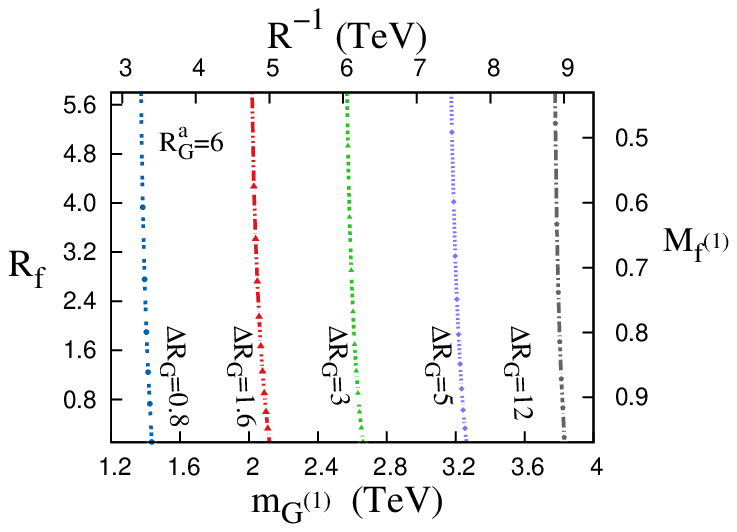}
\includegraphics[scale=1,angle=0]{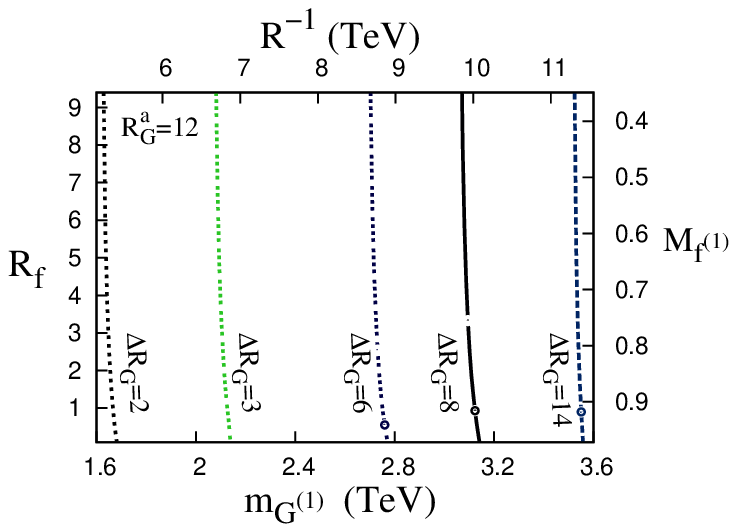}
\caption{Using the data of non-observation of a resonant production channel of $t\bar{t}$ signal at the LHC running at 13 TeV, we have shown excluded/allowed regions at 95\% C.L. in the $m_{G^{(1)}}-R_f$ plane for several choices of $\Delta  R_G = (r^b_G-r^a_G)R^{-1}$. Each panel specified by a particular value of $R^a_G$. The region to the left of a given curve is excluded by the CMS data \cite{Sirunyan:2017uhk}. $R^{-1}$ and $M_{f^{(1)}}(= m_{f^{(1)}}R)$ are shown in the upper and right-side axes respectively. \textquotedblleft G\textquotedblright stands for gluon.}
\label{ttd}
\end{center}
\end{figure}

In this section we have shown the excluded region of parameter space of the nmUED model utilising $t\bar{t}$ resonance production data of accumulated luminosity  2.6 $fb^{-1}$ reported by the CMS Collaborations \cite{Sirunyan:2017uhk}.\\\\
{\bf Case 1: BLKTs are present at {\boldmath$y=0$} and {\boldmath$y=\pi R$}}\\\\ 
In this case we have considered the BLKTs for fermions are equal at the two fixed boundary points, while KK-parity is broken by the unequal strengths of the gluon BLKTs. In Fig.\;\ref{ttd} there are three panels corresponding to three different values of $R_G^a$. In a particular panel there are several curves corresponding to different values of $\Delta R_G$. For a specific value of $R^a_G$ there is one-to-one correspondence of $m_{G^{(1)}}$  with $R^{-1}$ which is shown on the upper axis of the panels, as the KK-mass is mildly dependent upon $\Delta R_G$. Furthermore, one can determine $m_{f^{(1)}}$ using $M_{f^{(1)}} = m_{f^{(1)}} R$ corresponding to a particular value of $R_f$ and is displayed on the right-side axis. 

In each panel of Fig.\;\ref{ttd} the left portion of a curve specified by a fixed value of $\Delta R_G$ in the $m_{G^{(1)}}-R_f$ plane is excluded by the CMS data \cite{Sirunyan:2017uhk} at 95\% C.L.. The exclusion plots presented in Fig.\;\ref{ttd}
can easily be understood by conjunction of the two plots given in Figs.\;\ref{mass} and \ref{coup_plots}. From the Fig.\;\ref{coup_plots}-(a) it is seen that KK-parity non-conserving coupling is almost insensitive to $R_f$ while increases with $\Delta R_G$. Therefore, the resonance production cross section effectively depends on $R^a_G$ and $\Delta R_G$. Again from the Fig.\;\ref{mass}-(a) it is clear that for a fixed value of $R^a_G$ the KK-mass decreases with increasing values of $\Delta R_G$. One can check from the CMS data, that the cross section also decreases with increasing resonance mass. Therefore in order to match the observed data we need larger values of KK-parity non-conserving coupling due to the increment of $m_{G^{(1)}}$. Hence we need larger values of $\Delta R_G$ for a fixed values of $R_f$. Now if we take a larger value of $R^a_G$ we need higher value of $R^{-1}$ than that of by which we obtain the same resonance mass for lower values of $R^a_G$. At the same time higher values of $R^a_G$ brings the higher values of $R_f$. One should note that for any fixed value of $R_G^a$ (any one panel) the entire range of the mass given in the data cannot be covered. This happens, in order to match with the data, the KK-parity non-conserving coupling which requires for the model prediction of cross section for a particular $m_{G^{(1)}}$ varies only over a restricted range. These features are consistent with results in Fig.\;\ref{ttd}.

To this end, we are in a stage where we can discuss about the deviation of the limits of the model parameters by comparing the results obtained from current 13 TeV LHC analysis with previous 8 TeV LHC analysis \cite{drs}. But before going into that, let us spend some time to discuss the behaviour of resonance production cross section at 13 TeV LHC \cite{Sirunyan:2017uhk}. In this case the resonance production cross section for a particular mass is larger with respect to previous 8 TeV results \cite{cms8T}. One can explain this phenomena in the following way. For a resonance production the typical $x$ (given in Eq.\;\ref{int_sig}) values which we are probing are of the order of $\tau(\equiv m^2_{t\bar{t}}/S)$. A higher $S$ implies lower $\tau$, thus it not only increases range of $x$ integration, but also includes those region of $x$ for which parton density functions are higher in magnitude\footnote{A similar explanation is also valid for $\ell^+ \ell^-$ resonance production at the LHC which will be discussed in the next section. In this case also, for a particular resonance mass the production cross section for the $\ell^+ \ell^-$ resonance signal is larger with respect to previous 8 TeV data \cite{atlas8T}.}. 

If we translate the above phenomena in a particular panel specified by a particular value of $R^a_G$, then we can see that a smaller strength of KK-parity non-conserving coupling is sufficient to match the model prediction with new 13 TeV LHC data. Now KK-parity non-conserving coupling diminishes if we decrease the values of $\Delta R_G$ which in turn enhance the KK-mass (in this case $t\bar{t}$ resonance mass). Thus in this situation to obtain a specific value of resonance mass we need smaller value of $R^{-1}$ than that of 8 TeV analysis. For example if we consider the $t\bar{t}$ resonance mass 1.4 TeV for $R^a_G=3$, the values of other model parameters are, $\Delta R_G = 0.5$, $R^{-1}=2.5$ TeV and $R_f=1.53$ (corresponding fermion mass at first KK-excitation is 1.89 TeV) for $\sqrt{S}= 13$ TeV. However, when $\sqrt{S}=8$ TeV, in the same set of conditions the values of $\Delta R_G = 0.6$, $R^{-1}=2.6$ and $R_f=1.61$ (corresponding fermion mass at first KK-excitation was 1.95 TeV) \cite{drs}. Consequently the exclusion plots corresponding to different values of $\Delta R_G$ have been shifted towards the higher mass (resonance mass) region with respect to 8 TeV analysis.  Thus, in the current article we probe the higher mass region with smaller values of BLKT parameters (see Fig.\;\ref{ttd}). Further, the mass gap between $m_{f^{(1)}}$ and $m_{G^{(1)}}$ decreases as the value of $\sqrt{S}$ increases. This characteristics is also valid for other panels specified by different $R^a_G$.\\\\

\begin{figure}[t]
\begin{center}
\includegraphics[scale=1.5,angle=0]{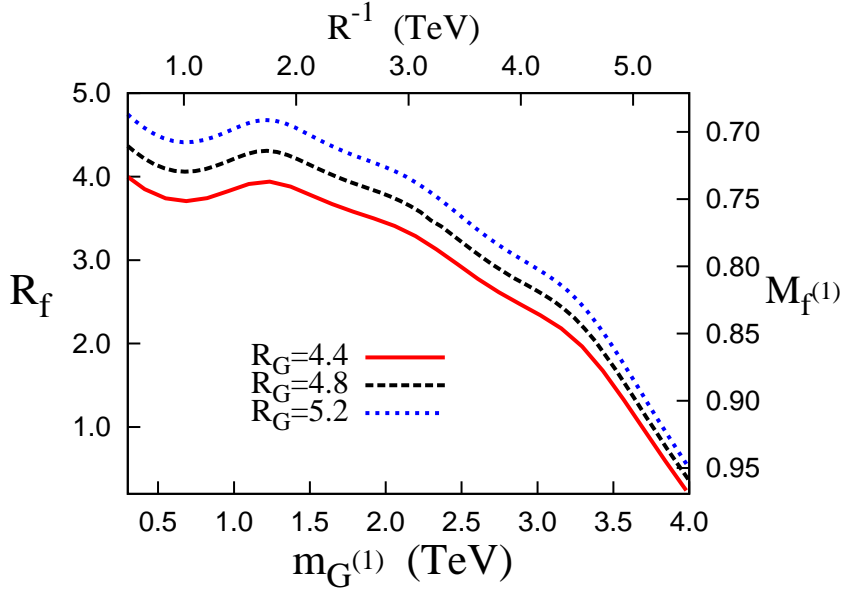}
\caption{Utilising the data of non-observation of a resonant $t\bar{t}$ signal at the LHC running at 13 TeV, we have shown the exclusion plots at 95\% C.L. in the  $m_{G^{(1)}}-R_f$ plane for several choices of $R_G$. The region below a particular curve is ruled out from the non-observation of a resonant $t\bar{t}$ signal in the 13 TeV run of LHC by CMS data \cite{Sirunyan:2017uhk}. $R^{-1}$ and $M_{f^{(1)}}(= m_{f^{(1)}}R)$ are shown in the upper and right-side axes respectively.\textquotedblleft G\textquotedblright stands for gluon.}
\label{tts}
\end{center}
\end{figure}

{\bf Case 2: BLKTs are present only at {\boldmath$y=0$}}\\\\
Now let us consider the case when fermion and gluon BLKTs are non-vanishing at only one fixed boundary point ($y=0$). In Fig.\;\ref{tts} we show the exclusion limits obtained by the 13 TeV results of CMS collaboration \cite{Sirunyan:2017uhk} for $t\bar{t}$ resonance search. Here we have shown the exclusion plots in the $m_{G^{(1)}}-R_f$ plane
for different values of $R_G$. The lower portion of a curve specified by $R_G$ has been excluded by the 13 TeV LHC data.

We can explain the exclusion plots on the basis of bottom of the left panels of the Figs.\;\ref{mass}  and \ref{coup_plots}.  It is quite evident form the Fig.\;\ref{mass}-(c), $M_{(1)}(= m_{G^{(1)}} R)$ has mild dependence on the value of
$R_G$. So, approximately one can take the mass of
$G^1$ to be simply proportional to $R^{-1}$. The  values of $R^{-1}$ are
displayed  in the upper axes of the panels in Fig.\;\ref{tts}.  For any $m_{G^{(1)}}$ 
the CMS data provides a limit for the
corresponding cross section times branching ratio. If we set the mass as fixed quantity, the experimental bound can be obtained by a specific
value for the KK-parity non-conserving coupling, which is a function $R_G$ and $R_f$. Alternatively, this can be viewed in the following way. As with increasing values of  $m_{G^{(1)}}$ the production cross section decreases, so to compensate this the KK-parity non-conserving coupling should be increased. At this point if we look at the Fig.\;\ref{coup_plots}-(c), then we can see that KK-parity non-conserving coupling shows oscillatory nature. However, we are interested in a region where $G^{1}$ is LKP, i.e., $m_{f^{(1)}} > m_{G^{(1)}}$, which belongs from a region where $R_G > R_f$. In this region KK-parity coupling increases with the increasing values of $R_G$. This is reflected in Fig.\;\ref{tts}.

Now in this single brane set up we would like to point out the departure of the limits of the model parameters that obtained from the current 13 TeV analysis with respect to 8 TeV analysis \cite{drs}. We have already discussed that KK-parity non-conserving coupling of lower strength is sufficient to match the model prediction with the 13 TeV LHC data \cite{Sirunyan:2017uhk}. Thus in the region where $R_G > R_f$, we can obtain lower strength of KK-parity non-conserving coupling by smaller values of $R_G$. This can easily probe the new 13 TeV LHC data. Let us consider the value of $t\bar{t}$ resonance mass at 1 TeV, the corresponding cross sectional value can easily be matched by lower values of $R_G$, e.g., 4.4 or 4.8. However, in the case of 8 TeV analysis the same set of conditions demand higher values of $R_G$, e.g., 5.0 \cite{drs}. As an artifact the values of $R_f$ have also been reduced in the 13 TeV analysis. For example, in the case of 13 TeV analysis the values of $R_f$ corresponding to the above mentioned values of $R_G$ are 3.84 and 4.20 respectively while in the case of 8 TeV analysis the value of $R_f$ was 4.5. This result revealed that the lower limits on BLKT parameters have been reduced due to higher centre of momentum energy ($\sqrt{S}$).

\section{{\boldmath$\ell^{+}\ell^{-}$} resonance search at 13 TeV and relative study with 8 TeV analysis} 
\vspace*{-.15cm}
Exactly in the same way as that of the $t\bar{t}$ resonance search, we have examined another signal at the LHC using the virtue of KK-parity non-conservation. In this case we have calculated the (resonance) production cross section of $B^1 (W^1_3)$ in $pp$ collisions at the LHC and their subsequent decay to $e^+ e^-$ and $\mu^+ \mu^-$, assuming $B^1 (W^1_3)$ to be the lightest KK-particles. We can exclude some portion of parameter space of this nmUED model using 13 TeV LHC data of integrated luminosity 36.1 $fb^{-1}$ of ATLAS collaboration \cite{Aaboud:2017buh} for $\ell^{+}\ell^{-} (\ell\equiv e, \mu)$ resonance production.\\\\
\hspace*{-.12cm}\vspace*{-.2cm}
{\bf Case 1: BLKTs are present at {\boldmath$y=0$} and {\boldmath$y=\pi R$}}\\

\begin{figure}[H]
\begin{center}
\includegraphics[scale=1, angle=0]{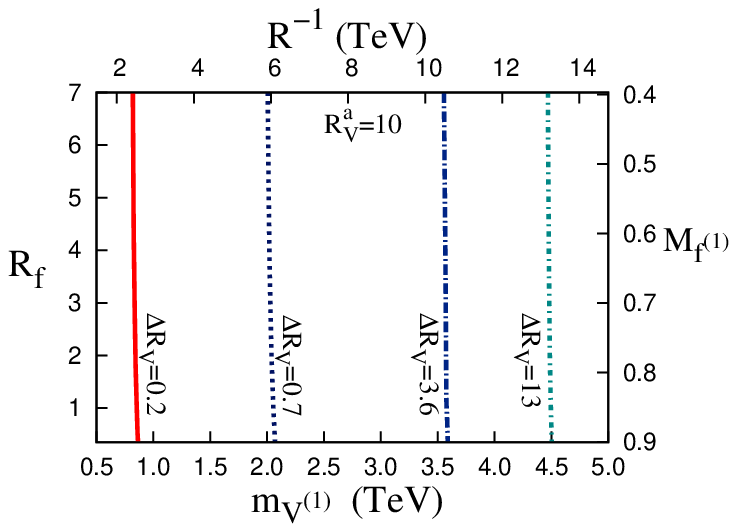}
\includegraphics[scale=1, angle=0]{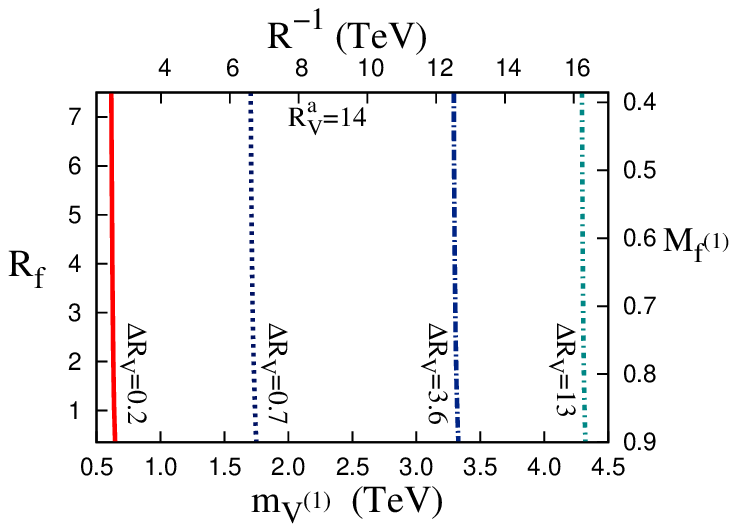}
\caption{Using the non-observation of a resonant $\ell^{+}\ell^{-}$ signal at the LHC running at 13 TeV, we have shown excluded/allowed regions at 95\% C.L. in the $m_{V^{(1)}} - R_f$ plane for different choices of $\Delta R_V= (r^b_V-r^a_V)R^{-1}$. Each panel specified by a particular value of
$R^a_V$. The region to the left of a given curve is excluded by the ATLAS data \cite{Aaboud:2017buh}. $R^{-1}$ and $M_{f^{(1)}} (=
m_{f^{(1)}} R)$ are shown in the upper and right-side axes
respectively. \textquotedblleft V\textquotedblright stands for $B/W_3$.}
\label{lld}
\end{center}
\end{figure}

Fig.\;\ref{lld} represents the case when the strengths of BLKTs for fermions are equal at both the boundary points however the strengths of BLKTs for electroweak gauge boson are unequal. Here we have shown the region excluded by 13 TeV LHC data \cite{Aaboud:2017buh} in two panels specified by different values of $R^a_V$. In the $m_{V^{(1)}}-R$ plane, the left region of a given curve specified by $\Delta R_V$ has been excluded by LHC data \cite{Aaboud:2017buh}. For any displayed value of $R_f$ we can measure the corresponding mass for the first KK-excitation of fermion using $M_{f^{(1)}} = m_{f^{(1)}} R$ ( plotted on right-side axis). The relevant values of $R^{-1}$  have been plotted on upper-side axis.

As we have already mentioned that for a chosen value of $\Delta R_V$ and $R^a_V$ the KK-parity-non-conserving couplings are almost independent of $R_f$. Thus $R_f$ has no governance on the production of $e^{+}e^{-}$/$\mu^{+}\mu^{-}$. Consequently signal rate is almost driven by $R^a_V$ and $\Delta R_V$. Furthermore, nature of the exclusion plots are very similar to the case of $t\bar{t}$ resonance signal. Therefore, following the same explanations (given for Fig.\;\ref{ttd}) for $t\bar{t}$ signal one can easily understand the exclusion plots of $\ell^{+}\ell^{-}$ resonance signal with the help of two figures Fig.\;\ref{mass}-(b) and Fig.\;\ref{coup_plots}-(b).

Despite the similar behaviour of the exclusion plots of the two different signals (for double brane set up) at the LHC, a unique feature has been observed in the exclusion plots (Fig.\;\ref{lld}) for $\ell^{+}\ell^{-}$ resonance signal. In the case the model prediction has been able to cover the entire LHC data of our concern. The reason is that the KK-parity non-conserving coupling needed for model prediction of cross section for a particular $m_{V^{(1)}}$ match with the data, varies over the entire range.

Let us see the deviations of the limits of the model parameters achieved from the 13 TeV analysis with that of obtained from 8 TeV analysis \cite{as}. To match the model prediction with the new 13 TeV LHC data \cite{Aaboud:2017buh} we require lower strength of KK-parity non-conserving coupling.  One can achieve this by reducing the value of $\Delta R_V$. However, it enhance the KK-mass (in this case $\ell^{+}\ell^{-}$ resonance mass). Therefore, to generate a typical value of resonance mass we need lower value of $R^{-1}$ than that of 8 TeV analysis. For example if we choose the $\ell^{+}\ell^{-}$ resonance mass as 2 TeV for $R^a_V=10$, the values of $\Delta R_V = 0.7$, $R^{-1}=5.9$ TeV and $R_f=6.64$. The corresponding fermion mass at first KK-excitation is 2.50 TeV. However, in the case of 8 TeV analysis, under the same set of conditions the values were $\Delta R_V = 1.5$, $R^{-1}=7.9$ and $R_f=6.70$ (corresponding fermion mass at first KK-excitation was 3.32 TeV) \cite{as}. Hence in the case of 13 TeV analysis the exclusion curves specified by different values of $\Delta R_V$ have been shifted towards the higher mass (resonance mass) region with respect to 8 TeV analysis.  Therefore, analysis with higher centre of momentum energy probe the higher mass region with relatively lower values of BLKT parameters (see Fig.\;\ref{lld}). In this case also the mass gap between $m_{f^{(1)}}$ and $m_{V^{(1)}}$ diminishes with increasing values of $\sqrt{S}$.\\\\

{\bf Case 2: BLKTs are present only at {\boldmath$y=0$}}

\begin{figure}[t]
\begin{center}
\includegraphics[scale=1.5,angle=0]{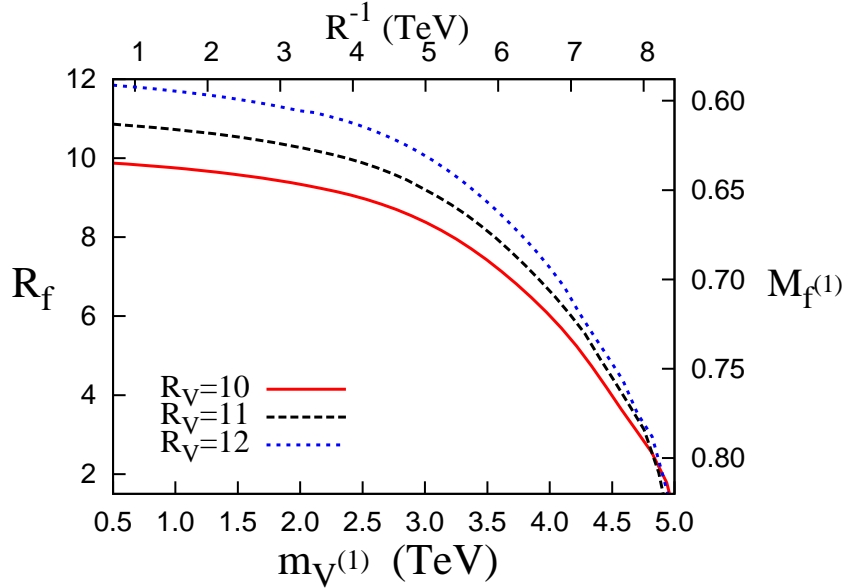}
\caption{Utilising the non-observation of a resonant $\ell^{+}\ell^{-}$ signal at the LHC running at 13 TeV, we have shown the exclusion plots at 95\% C.L. in the  $m_{V^{(1)}}-R_f$ plane for several choices of $R_V$. The region below a particular curve is ruled out from the non-observation of a resonant $\ell^{+}\ell^{-}$ signal in the 13 TeV run of LHC by ATLAS
 data \cite{Aaboud:2017buh}. $R^{-1}$ and $M_{f^{(1)}} (=
m_{f^{(1)}} R)$ are shown in the upper and right-side axes respectively.\textquotedblleft V\textquotedblright stands for $B/W_3$.}
\label{lls} 
\end{center}
\end{figure}

Let us study the case when fermion and electroweak gauge boson BLKTs are present at only one special  boundary point ($y=0$). In Fig.\;\ref{lls} we have plotted the exclusion curves in the $m_{V^{(1)}}-R_f$ plane for different choices of $R_V$. The lower portion of a curve has been disfavoured by the 13 TeV LHC data \cite{Aaboud:2017buh}.

In this case one can also see that the exclusion plots presented in Fig.\;\ref{lls} possess the same behaviour as in the respective case of $t\bar{t}$ signal. Therefore, on the basis of two figures Fig.\;\ref{mass}-(d) and Fig.\;\ref{coup_plots}-(d) it would not be difficult to understand the exclusion plots given in the Fig.\;\ref{lls}. It is quite evident from the Fig.\;\ref{mass}-(d), $M_{(1)}(= m_{V^{(1)}} R)$ has mild dependence on $R_V$. So one can take the mass of $V^1 (\equiv B^{1}, W^{1}_3)$ to be nearly proportional to $R^{-1}$ (the relevant values of $R^{-1}$ are displayed in the upper axis of the Fig.\;\ref{lls}). Here we can also find the KK-fermion mass of first excitation in a correlated way (using $M_{f^{(1)}} = m_{f^{(1)}} R$) from the right-side axis of this plot.

Now we are going to discuss the difference between the limits of the model parameters obtained from the current 13 TeV analysis and the 8 TeV analysis \cite{as}. We are particularly interested in a region where $V^1$ is LKP, for which we require $R_V > R_f$. If we see the Fig.\;\ref{coup_plots}-(d) then it will be understood that the region where $R_V > R_f$, the driving KK-parity non-conserving coupling decreases with the increasing values of $R_f$. As 13 TeV analysis demands lower strength of KK-parity non-conserving coupling to match the experimental data, one can achieve the lower strength of KK-parity non-conserving coupling by increasing the values of $R_f$. This is exactly reflected in the current analysis. For example we set the $\ell^{+}\ell^{-}$ resonance mass at 1.5 TeV, and set the value of $R_V$ at 11. The corresponding value of $R_f$ is 10.537 for which the value of fermion mass ($m_{f^{(1)}}$) of first KK-excitation is 1.5024 TeV. For 8 TeV analysis in the same set of conditions the value of $R_f$ was 10.239, and the corresponding value of $m_{f^{(1)}}$ was 1.5072 \cite{as}. This result provide the information that if we increase the value of centre of momentum energy, the value of $R_f$ increases, which in turn decreases the mass gap between $m_{f^{(1)}}$ and $m_{V^{(1)}}$. 
 
\vspace*{-.35cm}
\section{Conclusions}

We have studied the phenomenology of
KK-parity non-conserving UED model where all the
SM fields are allowed to propagate in $4+1$ dimensional space-time. We have
produced the non-conservation of KK-parity  by adding boundary localised terms of unequal strengths at the two boundary points of $S^1/Z_2$ orbifold. These boundary localised terms can be identified as cutoff dependent log divergent radiative corrections which play a crucial role to remove mass degeneracy in the KK-mass spectrum of
the effective 3+1 dimensional theory. 

In this paradigm we have produced KK-parity non-conservation in two different ways. In the first case we choose equal strengths of boundary terms for fermions at the two fixed boundary points ($y=0$ and $y=\pi R$) and parametrised by $r_f$, while we have considered unequal strengths of boundary terms $(r_V^a \neq r_V^b)$ for gauge boson. In the other set up we have considered non-vanishing boundary terms for fermion and gauge boson  only at the fixed point $y = 0$. The driving KK-parity non-conserving coupling vanishes in the $\Delta R_V = 0$ limit in the first case and
for $R_f = R_V$ in the later case. In these set up we calculate {\it two different} types of signal rate at the 13 TeV LHC. Specifically production of first KK-excitation of gluon ($G^1$) and its decay to $t\bar{t}$ pair. In the other case we study the production of first KK-excitation of neutral electroweak gauge bosons ($B^1~{\rm and}~W^1_3$) and their decay to $e^+ e^-$ and $\mu^+ \mu^-$ pair. Both the production and decay are the direct consequences of KK-parity-non-conservation. We compare our model predictions with the $t\bar{t}$ and $\ell^{+}\ell^{-} (\ell \equiv e, \mu)$ resonance production rate at the LHC running at 13 TeV $pp$ centre of momentum energy. The lack of observation of these signals at the LHC excludes a finite portion of the parameter space of this model.

Furthermore, in this article we have compared the limits obtained from the current 13 TeV analysis with that of previous 8 TeV analysis. With the increasing value of centre of momentum energy the parton density function with higher amplitude is folded in the production cross section.  Hence relatively lower strength of KK-parity non-conserving coupling is adequate to match the model prediction with the experimental data. However, in the case of 8 TeV analysis the required strength of KK-parity non-conserving coupling was higher with respect to present situation. Thus the change in relative strength of KK-parity non-conserving coupling has far reaching consequences in the model parameters. In both scenario for both signals the values of parameters have been changed significantly. In the case of double brane set up, for both the signals the asymmetric parameter ($\Delta R_V$) shifted towards the lower values, which immediately push the values of $R^{-1}$ towards the lower side. Apart from this the limits of fermionic BLT parameter ($R_f$) also decrease with that of 8 TeV analysis. Therefore, the mass gap between the resonance mass and the mass of first KK-excitation of the corresponding fermions has been diminished. In the case of single brane set up, 13 TeV analysis of $t\bar{t}$ resonance production decrease the values of $R_G$ with respect to 8 TeV analysis. However, the $\ell^{+}\ell^{-}$ resonance production increase the values of $R_f$ compare to 8 TeV analysis.

{\bf Acknowledgements} The author thanks Anindya Datta for many useful discussions.

\end{document}